\documentclass[runningheads]{llncs}
\usepackage[T1]{fontenc}
\usepackage{graphicx}

\usepackage{makecell}
\usepackage{rotating}
\usepackage{acronym}
\usepackage{textcomp}
\usepackage[table, x11names]{xcolor}
\usepackage{siunitx}
\usepackage{booktabs}
\usepackage{tabularx}
\usepackage{array}
\begin{document}
\title{A history of GDPR cookie banner compliance:\\ the roles of publishers, regulators and CMPs}
\titlerunning{A history of GDPR cookie banner compliance}
%
\author{Yana Dimova\inst{1}\orcidID{0000-0001-6558-2062} \and
Vincent Toubiana \inst{2}\orcidID{0009-0004-9968-1447} \and
Tom Van Goethem \inst{1}\orcidID{0000-0001-6846-9081} \and
Lieven Desmet \inst{1}\orcidID{0000-0001-5155-7472} \and
Wouter Joosen \inst{1}\orcidID{0000-0002-7710-5092}}
\authorrunning{Y. Dimova et al.}
%
\institute{DistriNet, KU Leuven, Belgium
\email{\{firstname.lastname\}@kuleuven.be} \and
CNIL, France
\email{vtoubiana@cnil.fr}}
\maketitle              
\begin{abstract}
	Since the introduction of the GDPR in 2018, cookie banners have become the primary mechanism for users to express preferences on online tracking and advertising. Consequently, their visual design and the options they present significantly influence user choice. Over time, the cookie banner landscape has evolved under the influence of key players, including publishers (website owners), regulators, and Consent Management Platforms (CMPs). 
	This paper presents an in-depth analysis of the roles of these three key actors and an examination of their impact on cookie banners' design and implementation within the context of EU law.
	Our results, based on a historical evaluation of 11,364 websites across 30 countries, indicate a positive evolution in the privacy landscape, with the compliance rate for websites featuring a ``reject all'' button increasing from 2.94\% in 2018 to 30.66\% in 2024.
	We analyze Data Protection Authority (DPA) activity and find a clear correlation between higher compliance rates and stronger regulatory action and guidance.
	Our experiments further show that compliance improvements are primarily driven by website owners, with CMPs showing little response to regulatory action or (indirect) influence on compliance rates.
	Our findings highlight the importance of more uniform collaboration and guidance among EU-level regulators to reduce interpretive divergence and simplify  cookie banner compliance, as well as the need for regulatory oversight of CMPs, which in turn could significantly enhance privacy for many websites and users. 
	Our work provides a foundation for academics, regulators, and industry to develop more effective strategies to motivate key players and promote greater user privacy.

\end{abstract}
\section{Introduction}

The General Data Protection Regulation (GDPR), which came into effect in May 2018, established a more formal definition of online consent. This had a significant impact on the ePrivacy Directive, which requires publishers to obtain valid consent before placing cookies on users' devices. Consequently, ``Cookie Banners'' have become the most visible manifestation of this new consent requirement, prompting publishers to adopt them on any website visited by EU data subjects.

While prior work has studied the positive changes due to the GDPR \cite{degeling2018we}, it has also highlighted the persistence of non-compliant interfaces \cite{bollinger2022automating,bouhoula2024automated} and their variation across geographical regions \cite{matte2020cookie,nouwens2025cross}. These non-compliant banners often include manipulative designs, that subtly nudge users toward less privacy-friendly choices \cite{nouwens2020dark,kirkman2023darkdialogs}.
While problematic design of cookie banners has been studied by prior work, less focus has been put on motivation behind compliance. In fact, DPAs have played a crucial role in addressing such issues of non-compliance. 

Their function is twofold: issuing guidance to clarify legal ambiguities and executing enforcement actions. For example, in December 2021, the French DPA (CNIL) sanctioned Google and Meta for failing to provide a ``reject'' button on the same layer as the ``accept'' option \cite{cnil_facebook_case_fr,cnil_google_case_fr}. Consequently, Google updated its banner design across the EU \cite{googlechangesbanner}, with Meta adopting similar changes outside of France \cite{dimova2022tracking}.

The CNIL case demonstrates that DPA enforcement exerts a direct extra-territorial impact. We examine the extent of these ``ripple effects'' across EU countries and websites, focusing on longitudinal regulatory influence. Our results reveal substantial variation in banner implementation, highlighting the pivotal role of DPA guidance and enforcement in driving compliance, and the need for technical standardization of consent.

Simultaneously, in response to the consent obligations under the GDPR, a new industry of specialized actors emerged: CMPs. Many publishers rely on CMPs to customize and deploy cookie banners on their website \cite{jha2024privacy,hils2020measuring}. 
Therefore, CMP design decisions have a broad influence on overall compliance.
We hypothesized that CMPs would respond to regulatory pressure and gradually converge toward a common set of GDPR‑compliant and privacy‑friendly cookie banner interfaces.
However, our results show the opposite.
An examination of consent interfaces from 18 CMPs shows substantial divergence, even among websites that only recently adopted a CMP. Among the five most widely used CMPs, for example, Cookiebot includes a reject button on 55.16\% of pages, whereas Cookies Consent does so on only 13.39\%.
While prior work focuses mainly on CMP-provided banners, our comparison of CMP and in‑house banners reveals that CMPs have contributed little to compliance improvements. Instead, most positive changes are driven by website owners who explicitly choose to implement them.


We analyze the historical evolution of cookie banners across 30 countries over a six-year period (2018–2024). To achieve this, we introduce a novel method for replaying cookie banners on archived web pages by combining data from two major web archives: the Wayback Machine and HTTPArchive. Our findings reveal a general positive trend toward greater privacy in cookie banner design over time. We integrate historical and cross-country perspectives, and dive deeper into the roles of three key actors (publishers, DPAs and CMPs) and their interactions. 

With our findings, we provide a better understanding of the motivations behind compliance and identify which actors have the greatest impact on the broader landscape. Our work can also help NGOs, plaintiffs, and regulators in establishing the most effective actions to increase compliance with regulation.

We summarize our main contributions: 
\begin{itemize}
	\item A longitudinal analysis demonstrating a general positive trend toward greater privacy in cookie banners over a six-year period.
    \item An examination of cross-country developments, highlighting the crucial role and positive impact of DPA regulatory actions.
    \item An investigation of cookie banners provided by 18 CMPs, which measures their limited influence on compliance and reveals that most privacy-positive changes are initiated by website owners.
\end{itemize}

\section{Related work}

\subsection{Web Archives}

The use of web archives offers significant advantages in research, notably in facilitating the reproducibility of web measurements. A comprehensive evaluation by Hantke et al. \cite{hantke2023you} of 13 public web archives found that the Internet Archive's Wayback Machine achieved the  best results in both suitability and performance. Prior work has leveraged web archives to investigate diverse topics, including the evolution of web tracking \cite{lerner2016internet}, changes in privacy policies \cite{degeling2018we,linden2018privacy,amos2021privacy}, the composition of the web \cite{agarwal2022way}, and the development of security features \cite{nikiforakis2012you,stock2017web,roth2020complex}. While related work by Jha et al. \cite{jha2024privacy} has examined the historical and geographical evolution of CMPs, no prior study, to our knowledge, has utilized web archives to specifically analyze cookie banners.

\subsection{Automated detection of cookie banners}

Our analysis relies on previously developed methods for automated cookie banner detection, a topic that has been extensively explored in prior work. A variety of approaches have been employed, including HTML- or CSS-based filtering of web page elements \cite{van2021impact,kirkman2023darkdialogs,rasaii2023exploring}, sometimes in combination with manual curation \cite{kampanos2021accept}. Other studies have utilized machine learning models on textual and visual elements \cite{khandelwal2022cookieenforcer} or a combination of multiple such methods \cite{gundelach2023cookiescanner,hausner2021dark,bouhoula2024automated}. For example, Bouhoula et al. \cite{bouhoula2024automated} present a tool that uses a combination of HTML/CSS filtering and machine learning to detect and interact with cookie banners, and classify cookies and declared purposes.

Many studies focus primarily on cookie notices provided by popular CMPs \cite{nouwens2020dark,habib2022okay,hils2020measuring,kirkman2023darkdialogs,bollinger2022automating} or supported by Interactive Advertising Bureau (IAB) Europe's Transparency and Consent Framework (TCF). Detection methods for TCF-enabled CMPs have included using the TCF consent signal \cite{matte2020cookie,rasaii2023exploring,zhang2024cschecker,nouwens2025cross}, public lists of CSS selectors \cite{jha2024privacy}, or methods that inspect HTTP requests \cite{habib2022okay,hils2020measuring,jha2024privacy} or page HTML \cite{nouwens2020dark,bollinger2022automating}, or a combination thereof \cite{kirkman2023darkdialogs}.

Unlike studies limited to manual analysis or popular CMPs, our research examines banners from 18 different CMPs alongside custom, in-house solutions. This broader scope provides a more comprehensive view of the cookie banner landscape.

\subsection{The impact of regulation on cookie banners}

Research on consent banners often addresses their interdisciplinary aspects. Warberg et al. \cite{warberg2023trends} examined the evolution of cookie banners on 911 news and media websites in the EU and US, observing an increase in reject options and a decrease in nudging strategies following GDPR enforcement. Similarly, Nouwens et al. \cite{nouwens2025cross} performed a cross-country analysis of the top 10,000 websites and found that only 15\% of consent interfaces were minimally GDPR compliant, with some country-specific differences. Other studies have focused on the evolution of CMPs \cite{jha2024privacy,hils2020measuring}.

Many studies have measured compliance rates across different countries within the EU \cite{nouwens2025cross,jha2024privacy,degeling2018we,trevisan20194,van2021impact,kampanos2021accept,libert2018changes}, or compared EU and US perspectives \cite{warberg2023trends,sanchez2019can,van2021impact}, and even considered an inter-continental scope \cite{rasaii2023exploring,fruchter2015variations}. Most studies offer only a static, single-point-in-time snapshot. Our work distinguishes itself by providing a historical, longitudinal analysis over a period of six years, focusing on the long-term impact of the GDPR.

The lawfulness of cookie banners has also been a major research focus. Studies have addressed issues such as implicit consent \cite{kretschmer2021cookie,degeling2018we,sanchez2019can,mehrnezhad2020cross,nouwens2020dark,bollinger2022automating,trevisan20194,matte2020cookie,zhang2024cschecker,rasaii2023exploring,sheil2022fianan,leenes2015taming,krisam2021dark,alharbi2023empirical,wesselkamp2021depth,libert2018changes,bouhoula2024automated} and the presence of dark patterns \cite{warberg2023trends,nouwens2020dark,habib2022okay,kirkman2023darkdialogs,sheil2022fianan,krisam2021dark,soe2020circumvention,alharbi2023empirical,bouhoula2024automated}, including pre-checked options \cite{nouwens2025cross,mehrnezhad2020cross,nouwens2020dark,matte2020cookie,zhang2024cschecker,kirkman2023darkdialogs,sheil2022fianan} and cookie walls \cite{kirkman2023darkdialogs,sheil2022fianan,leenes2015taming,alharbi2023empirical}. Large-scale measurement studies have exposed the  widespread non-compliance on cookie banners, finding that 72.2\% of websites violated at least one GDPR requirement \cite{bouhoula2024automated}.



\section{Historical evolution of cookie banners}
This section examines the historical evolution of cookie banners. We first outline our methodology for replaying archived webpage snapshots to automatically identify GDPR violations. We then analyze trends in banner prevalence and the evolution of user options over time.

\subsection{Methodology}
Our study examines the evolution of cookie banners across the EU from 2018 to 2024. Unlike prior research dependent on web crawling, we leverage two of the largest web archives to retrospectively and automatically analyze historical snapshots.

\subsubsection{Web Archives}
The HTTPArchive \cite{HTTPArchive} is a valuable open-source dataset that has been tracking the composition of the web since 2012. Each month, it performs crawls of millions of web pages (12 million URLs in the latest period). The resulting data is stored in HAR (HTTPArchive) format, a JSON-formatted file that logs a web browser's interactions with a website.
Since July 2018, HTTPArchive's crawls have been based on the Chrome User Experience Report (CrUX) \cite{crux}, a list of the most visited websites globally. The data, captured from US-based agents, includes all requests and responses for each web page and is stored on BigQuery. To the best of our knowledge, no historical web archives of comparable scale or completeness exist from an EU perspective. 

To ensure comprehensive capture of dynamic cookie banners, we supplemented HTTPArchive snapshots with data from the Wayback Machine \cite{waybackmachine}. This dual-archive approach is essential because HTTPArchive occasionally misses dynamic request chains required to load CMP banners. By using the Wayback Machine’s API to retrieve missing resources, we can accurately replay historical banners in their entirety.


\subsubsection{Replaying web pages}
We use a multi-tiered approach combining HTTPArchive and the Wayback Machine. First, we extract relevant snapshots from HTTPArchive and store them locally, then preprocess the data to address missing or incomplete response bodies, which are common due to HTTPArchive’s cutoff limits. For each resource, we check for missing bodies and, if found, attempt retrieval via the Wayback Machine’s API, ignoring images and fonts as they are unnecessary for banner layouts and omitted by HTTPArchive by default. This step is crucial for managing the Wayback Machine’s rate limit of 60 requests per minute \cite{ratelimitWM}. Without this pre-emptive approach, attempting to retrieve all missing resources on the fly would significantly slow down our crawler, with each page taking several minutes to replay.

To replay pre-completed snapshots, we instruct the browser to visit selected URLs through an HTTP proxy that intercepts traffic and redirects requests to archived snapshots. If a resource is missing, we fetch it from the Wayback Machine. If the resource is not available in both archives, we return an empty response. This ensures all interactions remain offline and prevents any connection to the live web.

\subsubsection{Website selection}

While past studies have looked at the immediate effects of the GDPR in May 2018, our work examined its long-term impact on websites in all 30 EEA countries (27 EU countries, Norway, Liechtenstein and Iceland).

To select our sample, we filtered the CrUX list by country code top-level domain (ccTLD) (e.g., \textit{.fr} for France). From this set, we identified websites with available snapshots spanning the entire analysis period. Finally, we randomly selected 500 websites per country. Because some countries had fewer than 500 websites available in CRuX (see Table 1), our final dataset consists of 11,364 websites.

\begin{table*}[!h]
	
	\resizebox{\textwidth}{!}{
		\begin{tabular}[width=\textwidth]{ c|c|c|c|c|c|c|c|c|c|c|c|c|c|c|c|c|c|c|c|c|c|c|c|c|c|c|c|c|c|c|c } 
			& \begin{sideways} Total \end{sideways} & \begin{sideways} Austria \end{sideways} & \begin{sideways} Belgium \end{sideways} & \begin{sideways} Bulgaria \end{sideways} & \begin{sideways} Croatia \end{sideways} & \begin{sideways} Cyprus \end{sideways} & \begin{sideways} Czech \end{sideways} & \begin{sideways} Denmark \end{sideways} & \begin{sideways} Estonia \end{sideways} & \begin{sideways} Finland \end{sideways} & \begin{sideways} France \end{sideways} & \begin{sideways} Germany \end{sideways} & \begin{sideways} Greece \end{sideways} & \begin{sideways} Hungary \end{sideways} & \begin{sideways} Ireland \end{sideways} & \begin{sideways} Italy \end{sideways} & \begin{sideways} Latvia \end{sideways} & \begin{sideways} Lithuania \end{sideways} & \begin{sideways} Luxembourg \end{sideways} & \begin{sideways} Malta \end{sideways} & \begin{sideways} Netherlands \end{sideways} & \begin{sideways} Poland \end{sideways} & \begin{sideways} Portugal \end{sideways} & \begin{sideways} Romania \end{sideways} & \begin{sideways} Slovakia \end{sideways} & \begin{sideways} Slovenia \end{sideways} & \begin{sideways} Spain \end{sideways} & \begin{sideways} Sweden \end{sideways} & \begin{sideways} Iceland \end{sideways} & \begin{sideways} Liechtenstein \end{sideways} & \begin{sideways} Norway \end{sideways} \\
			\hline
			\makecell{Number of \\ websites} & \textbf{11364} & 500 & 500 & 377 & 320 & 30 & 500 & 500 & 190 & 500 & 500 & 500 & 500 & 500 & 500 & 500 & 182 & 316 & 65 & 19 & 500 & 500 & 500 & 500 & 500 & 247 & 500 & 500 & 109 & 9 & 500\\
			\rowcolor{gray!20} \makecell{Max detected \\banners} & \textbf{4060} & 155 & 178 & 128 & 131 & 9 & 194 & 112 & 73 & 151 & 137 & 172 & 176 & 220 & 190 & 154 & 95 & 138 & 16 & 9 & 149 & 217 & 166 & 205 & 212 & 116 & 196 & 184 & 37 & 1 & 139  \\
			\makecell{Detected\\ banners} & \textbf{2479} & 99 & 100 & 85 & 94 & 4 & 103 & 66 & 24 & 52 & 81 & 92 & 111 & 157 & 114 & 106 & 50 & 92 & 8 & 7 & 93 & 163 & 111 & 131 & 124 & 96 & 148 & 109 & 20 & 1 & 38  \\
			
	\end{tabular}}
	\caption{The first row of this table shows the number of websites crawled for each country. The second row shows the maximum number of cookie banners detected (in September 2024). The third row shows the number of websites which include a cookie banner throughout the whole analysis period.}
	\label{ref:table1}
	
\end{table*}


We select country-code top-level domains (ccTLDs) for two reasons. First, this ensures the dataset represents popular websites within each target country. Second, although snapshots are captured via US-based IP addresses, ccTLDs target European audiences, as confirmed in the \textit{Pammer and Alpenhof} case \cite{pammer-alpenhof-judgement,guidelines-territorial-scope}. Therefore, these websites fall within the territorial scope of GDPR Article 3 \cite{gdpr}.

We performed several analyses to confirm this latter claim. First, we leveraged the country-specific CrUX popularity information, and compared the popularity of sites with an EEA ccTLD to its global popularity. We find that the vast majority (over 99\%) of sites  in our dataset are much more popular in the EEA than globally, clearly indicating that these target European citizens. We excluded a few exceptions, such as vanity domains (e.g., \texttt{linktr.ee}) or compromised websites.
Second, we confirmed that none of the 11,364 websites in our dataset redirect from an EU ccTLD to a global, non-EU domain (e.g., \texttt{youtu.be} to \texttt{youtube.com}), which would potentially complicate their GDPR obligations. Finally, our assessment of the impact of US-based access (detailed in Section \ref{validation}) shows that 91.6\% of cookie banners are captured consistently regardless of the access location.
Our methodology aligns with prior work that adopts a similar ccTLD-based approach \cite{jha2024privacy}. 

\subsubsection{Cookie banners analysis}

To detect and analyze cookie banners, we extended the automated tool developed by Bouhoula et al. \cite{bouhoula2024automated}. Built on the OpenWPM framework, this tool identifies and interacts with banners to report GDPR violations for both CMP and in-house solutions. We added support for the primary languages of all 30 target countries by integrating the Google Cloud Translate API to translate banner text before processing.

The framework locates candidate banners through HTML filtering and z-index inspection. It then validates these candidates using a language model to identify cookie-related terms, achieving 100\% precision and 86.9\% recall. A machine learning model classifies banner options (e.g., accept, reject, settings) and links them to specific cookie sets. Finally, cookies are categorized as either 1) analytics and advertising or 2) essential and session based on their purpose. For detecting missing ``Reject'' buttons, the tool yields 91.5\% precision and 65.2\% recall, meaning false positives are rare, though some non-compliant banners are missed. While our findings represent a lower-bound estimate, they remain suitable for identifying long-term trends.

\subsubsection{Detecting GDPR violations}
\label{sec:detecting_gdpr_violations}
A multitude of factors dictate compliance with data protection laws, ranging from the adequacy of user information to the deployment of pre-selected choices and the installation of tracking cookies without explicit consent. While comprehensive compliance involves complex, qualitative dimensions, this study focuses on the foundational and objective parameters of lawful consent. To mitigate inconsistencies arising from divergent national interpretations, such as varying regulations on ``cookie walls'' \cite{morel2022your}, we deliberately exclude metrics that require subjective legal judgment. By restricting our analysis to automatically detectable and consistently interpreted infractions, we ensure an unbiased assessment that facilitates reliable cross-country and longitudinal comparisons.

Within this methodological framework, we evaluate cookie banners that deploy tracking cookies and are thus legally mandated to provide an equally accessible rejection mechanism. The absence of a clear ``reject'' option violates the GDPR requirement for freely given consent, which dictates that declining consent must be as effortless as accepting it, a principle firmly upheld by the CNIL \cite{cnil_facebook_case_fr} and the EDPB taskforce \cite{cookie_banner_taskforce}. Consequently, we utilize this specific consent requirement as a primary proxy for broader GDPR compliance.

Furthermore, initializing tracking cookies without offering an immediate rejection mechanism violates the ePrivacy Directive (Art. 5.3) \cite{eprivacy,guidelinesscopeeprivacy}, which strictly requires prior consent for non-essential cookies. Because our data collection simulated access from outside the EU, we cannot definitively confirm actual ePrivacy violations within our dataset. Therefore, our findings should be interpreted as potential ePrivacy violations. This extrapolation remains highly robust, given that our comparative analysis between US and EU vantage points revealed no significant divergence in website behavior (detailed in Section 3.2).

\subsubsection{Crawls}

In the final step, we crawled 11,364 websites across 25 time periods and 30 countries using an automated detection tool interfaced with archived snapshots via a proxy. The primary bottleneck was live requests to the Wayback Machine API, which extended each crawl of 500 websites to approximately 36 hours. The process was deployed on a university cloud network using machines with 16 GB RAM and 4 CPUs.
Out of a total of 284,100 websites (25 time points multiplied by 11,364 websites), we successfully crawled 277,440 (97.66\%) of them.

\subsection{Web Archives: Benefits and Constraints}
\label{validation}

Web archives provide temporal stability by eliminating some of the disadvantages of live web crawls. However, they present limitations such as browser compatibility issues \cite{hantke2025web} and incomplete resource retrieval \cite{lerner2016internet}. Because some dynamic content may be replayed incorrectly, our cookie banner detection rates represent a lower-bound estimate. We quantify  this potential bias through a three-fold validation.

\paragraph{CMP Detection Comparison}
By matching HTTP requests to known CMP domains (Section \ref{sec:method_cmp_detection}), we found that a substantial share of banners remains uncaptured, increasing from 41.49\% in 2018 to 49.59\% in 2024. However, these missed banners are distributed across various providers (e.g., Cookiebot, OneTrust), suggesting our longitudinal trends remain representative of the broader ecosystem.

\paragraph{Manual Analysis}
We manually analyzed 500 websites to assess crawler errors, archive limitations, and geographical bias. Our automated methodology successfully captured 53.2\% of banners found manually. 

To test IP location bias, we crawled 500 websites with a EU IP address and a US IP address and compared the results. EU crawls detected 73.72\% of banners, while US crawls detected 64.96\%. This 8.4\% disparity is relatively low because we focused on country-code TLDs (ccTLDs), which, as Van Eijk et al. \cite{van2021impact} noted, often influence compliance more than the visitor's location, particularly when compared to generic \texttt{.com} domains.

\paragraph{Comparison with Prior Work}
Our September 2024 detection rate was 35.73\%. While Bouhoula et al. \cite{bouhoula2024automated} reported 57.16\% using the same crawler on live sites, our lower rate is expected given the archive and geolocation constraints. Our results align with the broad range found in existing literature, where automated detection rates vary from 24\% \cite{kirkman2023darkdialogs} to 67\% \cite{nouwens2025cross}. 
We conclude that our findings remain representative of the broader cookies banners landscape and their evolution of time, despite the acknowledged limitations.

\subsection{Results}

\begin{figure}[h!]
	\includegraphics[alt ={This figure shows the number of cookie banners that we detected over time, starting from 2479 websites with cookie banners and increasing to 4060 by 2024.}, width=\linewidth]{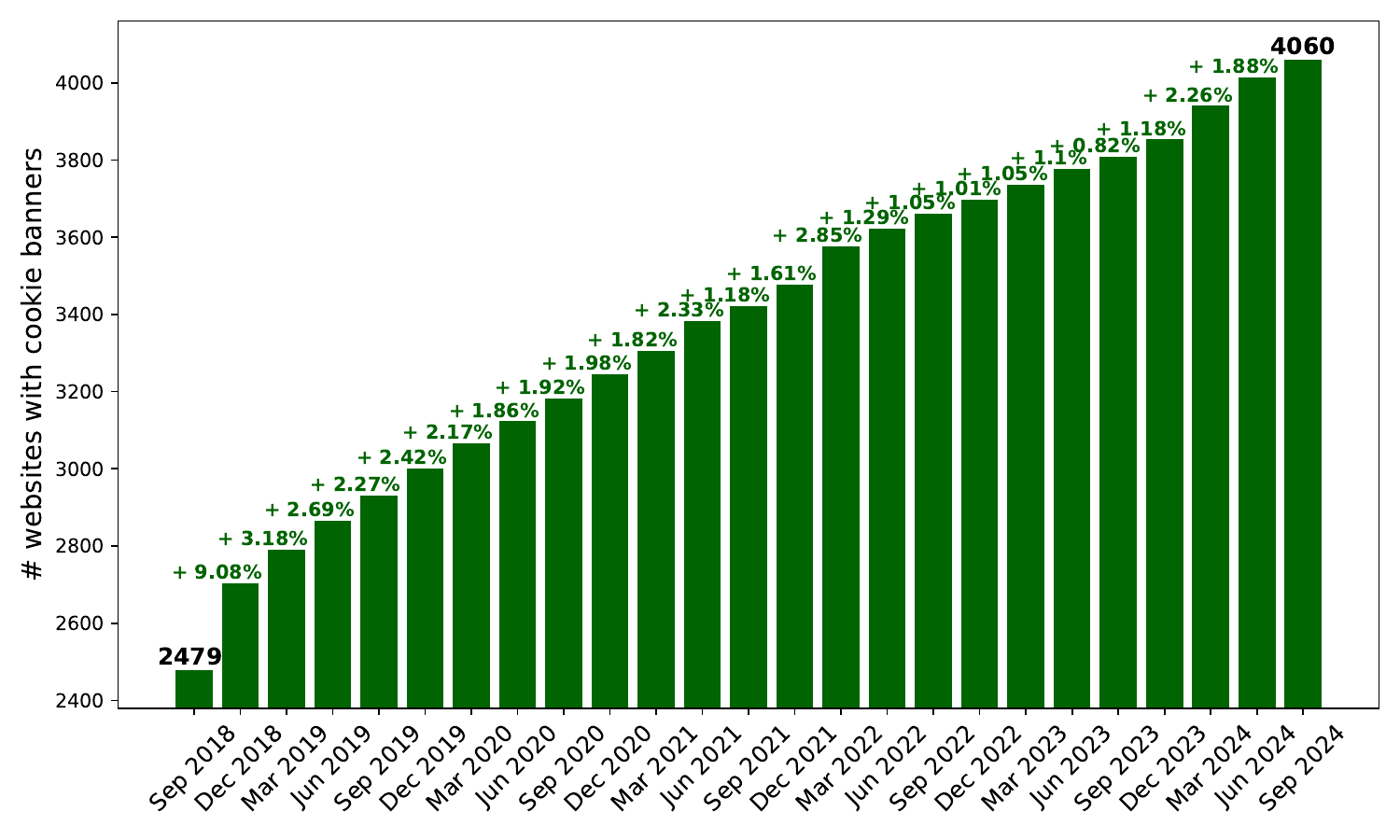}
    \caption{Number of detected cookie banners throughout time}
    \label{fig:number_banners}

\end{figure}

We crawled 11,364 websites and detected a cookie banner on 4,060 (35.73\%) of them. The number of cookie banners detected increased continually throughout the measurement period, as shows in Figure \ref{fig:number_banners}. The detection rate varied across EU countries, from 22.4\% in Denmark to 52.2\% in Latvia. A detailed country breakdown is shown in Table \ref{ref:table1}.
Our historical analysis over six years (September 2018–September 2024) shows a steady increase in cookie banner adoption. At the start, 2,479 (21.81\%) websites consistently included a banner. By September 2024, this number rose to 4,060 (35.73\%), a 63.78\% increase.
This growth aligns with the post-GDPR regulatory landscape. Following GDPR’s introduction, website owners likely became more aware of their legal obligations to obtain user consent for non-essential third-party services, such as analytics and tracking, prompting gradual adoption of cookie banners.

\subsubsection{Cookie banners options}
\label{sec:cookie_banners_options}
Next, we examine the evolution of consent options on cookie banners, focusing on websites that consistently displayed a banner throughout the analysis period. We categorize these into four types: 1) banners offering both ``accept'' and ``reject'' options, 2) banners with only an ''accept`` option, 3) banners with only a ``reject'' option, and 4) banners providing no choice at all.
 
\begin{figure}[htbp!] 
	\centering
	\makebox[\linewidth][c]{%
		\includegraphics[alt= {This figure presents the options given to users and how they evolve over time. We observe that banners with accept and reject gave increased from only 5\% of websites to being prevalent on about 30-35\% of websites in 2024. The evolution of accept-only and no options banners shows the opposite evolution: they were very prevalent in 2018 but have decreased in use by 2024 in favor of banners inlcluding accept and reject options.}, width=1.7\linewidth]{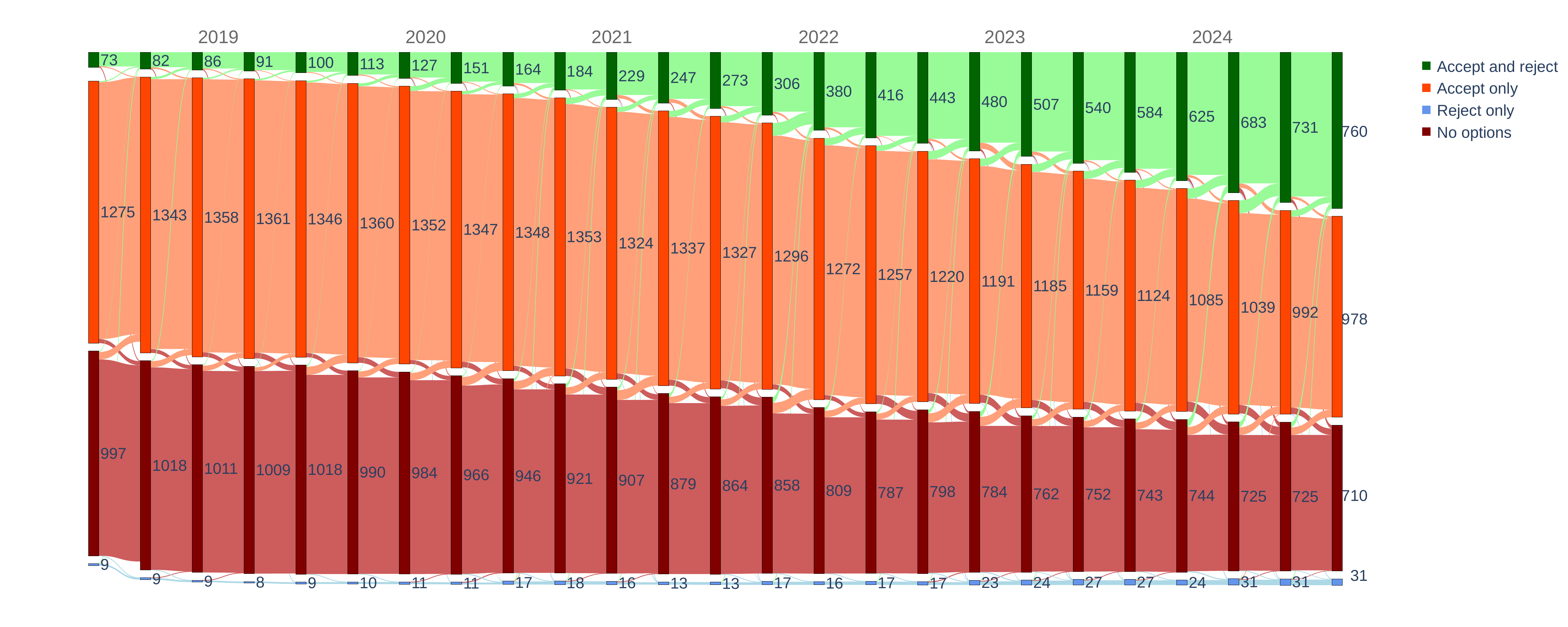}%
	}
	\caption{Number of cookie banners over time with ``Reject'' options, only ``Accept'' options, and no options.}
	\label{fig:cookie_banners_options}
\end{figure}

As shown in Figure \ref{fig:cookie_banners_options}, which illustrates the evolution of these four types, we observe several positive trends. First, banners offering both accept and reject options increased steadily over time, from 73 (2.94\%) websites in September 2018 to 760 (30.66\%) in September 2024. This reflects a positive shift toward providing users with more choice.

Second, we note a short-term rise in banners featuring only an ``Accept'' button between September 2018 and March 2019. This likely occurred as many websites initially implemented passive banners following the GDPR’s introduction. Subsequent regulatory guidance likely prompted website owners to recognize the unlawfulness of this practice, leading them to adapt banners to include active consent options. Following this period, the number of accept-only banners declined steadily in favor of banners that include a reject option.

\subsubsection{Periods with most change}
\begin{figure}[h!]
	\includegraphics[alt={The figure shows the relative growth of number of banner with accept and reject, with the greatest increase around 2022 with 7.5\% and another peak by the end of the time period with an increase of 8.82\% by 2024.}, width=\linewidth]{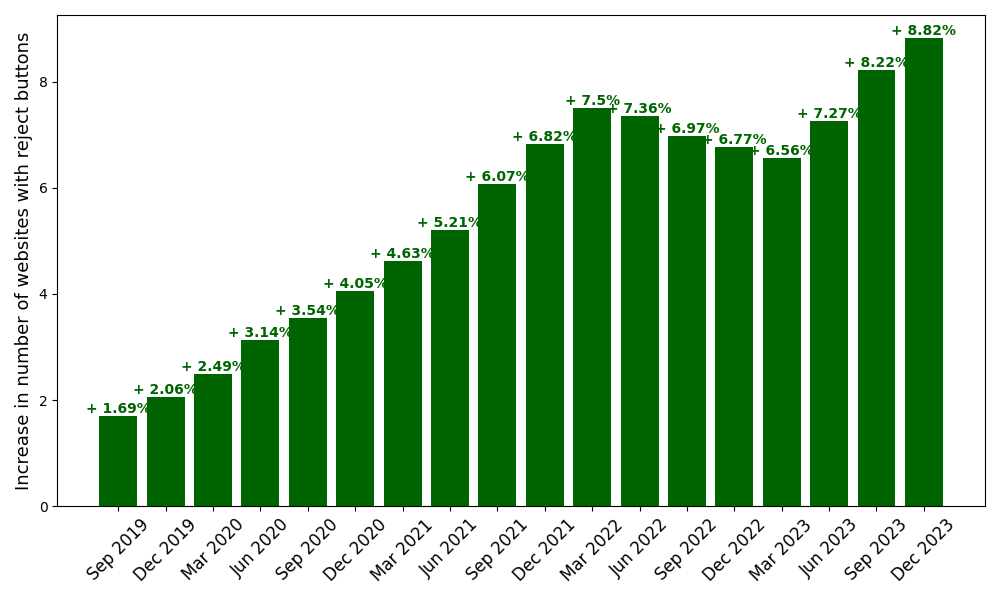}
    \caption{This figure shows the growth of ``Reject'' options relative to banners previously lacking this feature. We compare the growth in one year starting from the given period, with the year before. E.g., between September 2019 and June 2020  1.69\% more websites adopted cookie banners with accept and reject options as compared to Sep 2018-Jun 2019.}
    \label{fig:relative_growth_global}

\end{figure}

To identify during which period cookie banners became more privacy-focused, we calculated the growth rate of banners offering both ``accept'' and ``reject'' options relative to those that lacked them. These trends are detailed in Figure \ref{fig:relative_growth_global}.

We observed two major shifts. The first peak occurred between January and December 2022, when 7.5\% more websites began displaying reject buttons on their banners. The second period took place toward the end of our analysis, between December 2023 and September 2024, with an 8.82\% increase in banners offering a reject option.

It appears that a significant shift in the compliance of cookie banner interfaces occurred in 2022, followed by an even more impactful shift that began in early 2024 and continued until the end of our analysis period. While the main driver for this positive evolution is not yet clear, it is possible that this behavior will continue to propagate into late 2024 and 2025.

The changes in cookie banner features can be attributed to numerous factors. These include country-specific developments, such as regulatory actions, as well as EU-wide ripple effects (e.g., from cases of the European Court of Justice). Additionally, website owners often employ CMPs to manage their cookie banners, which can significantly influence how banners are displayed. We explore these considerations in Sections 5 and 6.

\section{Cross-country analysis}
In this section, we shift the perspective from EU-wide trends to country-specific developments, exploring which countries have experienced the most significant increase in cookie banner compliance over time.

\subsection{Methodology}

For the analysis, we rely on the same dataset, described in section 3.1. To assess how a website’s territorial scope influences its banner, we compare GDPR compliance across countries by focusing on banners without reject options (at any layer).

\subsubsection{Activity level of DPAs}

Our analysis focuses on two defining activities for DPA involvement: the imposition of sanctions and fines for GDPR infringements and the publication of regulatory guidance (e.g., guidelines, recommendations).

We extracted this information from noyb's GDPRhub \cite{gdprhub}, an open-source wiki summarizing over 2,000 national cases. We identified a total of 287 relevant cases by searching for the keyword ``cookie'' in their summaries. This allowed us to compile the number of cases and total fine amounts in euros for each country. Furthermore, we referenced noyb's ``Consent Banner Report'' \cite{NOYBreport} for an overview of all pertinent EU and national guidelines on cookie banners.

\subsubsection{Correlation test}
To investigate the relationship between DPA activity and compliance, we performed a Pearson correlation test. We tested the correlation between 1) compliance rates and the presence of regulatory guidance (as a binary parameter), and 2) compliance and enforcement (as measured by the number of cases).

\subsection{Results}

\subsubsection{Cross-country differences}

\begin{table}[!h]
\centering
\small 
\rowcolors{2}{gray!20}{white}
\begin{tabular*}{\columnwidth}{l|c|c|c|r|c}
    \toprule
    \textbf{Country} & \makecell{\textbf{Avg. banners}\\ \textbf{with Reject(\%)}} & \makecell{\textbf{Latest banners} \\ \textbf{with Reject (\%)}} & \makecell{\textbf{\# of}\\\textbf{Cases}} & \makecell{\textbf{Total Fines}\\\textbf{(EUR)}} & \makecell{\textbf{Guidelines}\\\textbf{Published}} \\
    \midrule
    Denmark & 35.15 & 60.61 & 15 & --- & 2020 \\
    France & 23.51 & 41.98 & 43 & \num{645250000} & 2020 \\
    Austria & 20.08 & 44.44 & 19 & --- & 2023 \\
    Germany & 19.61 & 38.04 & 25 & --- & 2022 \\
    Finland & 18.69 & 44.23 & 5 & \num{1100000} & 2022 \\
    Ireland & 18.24 & 40.35 & 4 & \num{225000000} & 2020 \\
    Italy & 17.66 & 37.74 & 21 & \num{51732546} & 2021 \\
    Netherlands & 16.43 & 25.81 & 14 & \num{1165500} & 2019 \\
    Belgium & 16.08 & 31.00 & 34 & \num{604140} & 2023 \\
    Slovakia & 15.36 & 42.74 & --- & --- & --- \\
    Spain & 14.78 & 43.24 & 74 & \num{37021600} & \makecell{2019, 2020, \\ 2023} \\
    Latvia & 14.40 & 30.00 & 2 & --- & 2022 \\
    Czech Rep. & 13.44 & 39.81 & 2 & \num{14160363} & unknown \\
    Croatia & 13.32 & 27.66 & 5 & \num{133000} & --- \\
    Portugal & 12.40 & 33.33 & 1 & --- & --- \\
    Slovenia & 11.38 & 17.71 & --- & --- & --- \\
    Greece & 9.95 & 20.72 & 4 & \num{440000} & 2020 \\
    Sweden & 9.91 & 26.61 & 9 & \num{6955136} & --- \\
    Hungary & 8.66 & 20.38 & 3 & \num{25000} & --- \\
    Romania & 8.18 & 19.08 & 5 & \num{55000} & --- \\
    Bulgaria & 6.73 & 14.12 & --- & --- & --- \\
    Poland & 5.89 & 20.25 & 2 & \num{40000} & --- \\
    Lithuania & 4.74 & 14.13 & --- & --- & --- \\
    \textbf{Average} & \textbf{14.55} & \textbf{31.91} & \textbf{12.5} & & \\
    \bottomrule
\end{tabular*}
\caption{Compliance rates differences across countries represented by the presence of ``Reject'' options. The second column shows the average compliance rate throughout the whole time period, while the third column refers to the compliance rate in September 2024. DPA activity is represented by the number of cases on cookies, total amount of fines and the publication of guidelines on cookie banners.}
\label{table:ar_banners_per_country}
\end{table}

Cookie banner compliance varied significantly across countries. Denmark maintained the highest average compliance (35.15\%), followed by France (23.51\%) and Austria (20.08\%). By 2024, the top performers were Denmark (60.61\%), Austria (44.44\%), and Finland (44.23\%).

While Spain (43.24\%) and Slovakia (42.74\%) reached the top five in 2024, their lower historical averages suggest a very recent surge in compliance. Conversely, Lithuania (4.74\%), Poland (5.89\%), and Bulgaria (6.73\%) consistently exhibited the lowest compliance rates.

\subsubsection{The impact of regulatory action}

DPA engagement varies significantly across the EEA. For instance, the Spanish DPA has processed the most cookie-related enforcement cases (74), totaling €37,021,600 in fines. Meanwhile, the French CNIL has levied the highest aggregate fines, totaling €645,250,000 across 43 cases.

Our findings suggest that DPA activity measurably impacts compliance. We identified a moderate positive correlation between compliance and regulatory action \texttt{(r=0.42,p=0.046)} and a strong positive correlation with the publication of regulatory guidance \texttt{(r=0.62,p=0.002)}. While other factors, such as a country's general privacy stance, reputational risk, and economic considerations, likely influence compliance, a full evaluation of these multidisciplinary aspects remains outside this study's scope.


\subsubsection{Case study: France}
\begin{figure}[h]
	\includegraphics[alt={In this figure we see that the increase in compliance on French websites around the time of the CNIL case is very high. We observe an increase of 21.69\% for January 2022, way higher than the overall compliance evolution.}, width=\linewidth]{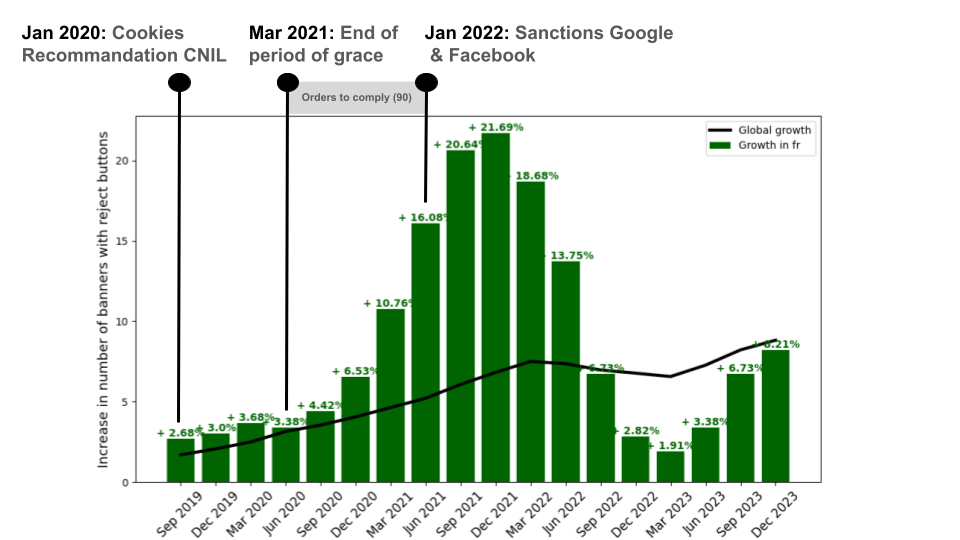}
	\caption{Increase in number of banners with reject buttons over time in France}
	\label{fig:evolution_france}
\end{figure}

In France, we observe a consistently high compliance rate throughout the analysis period. The CNIL case likely caused a significant increase in the number of reject buttons after January 2022. As shown in Figure \ref{fig:evolution_france}, we observe that in March 2022, 16.08\% more websites were compliant. The compliance rate continued to increase, with an additional 21.69\% of websites becoming compliant by September 2022. This phased increase could be attributed to some websites making changes immediately after the CNIL ruled Google's cookie banner non-compliant (on December 31st, 2021), while others only made changes months later, after Google announced its intent to redesign its banner according to CNIL's guidelines \cite{googlechangesbanner} (on April 21, 2022).

The case against Meta and Google appears to have had a significant impact in France. However, we did not observe a similar increase in compliance after the CNIL published a recommendation on the use of cookies \cite{recommendationCookiesCNIL} (in January 2020) or after the grace period for companies to comply with these guidelines ended (in March 2021) \cite{endperiodofgraceCNIL}. France is one of three countries where the evolution of the compliance rate mimics the global trend, peaking around 2022 and 2024 (the others being the Czech Republic and Slovakia).

\section{Cross-CMP analysis}
Website owners frequently rely on CMPs for cookie banner configuration. While CMPs offer default settings, owners often customize these by adding ``reject'' buttons, implementing dark patterns, or modifying button text. Consequently, CMPs directly shape banner design and user controls, significantly influencing the overall privacy-friendliness of a site. This section explores the evolving role of CMPs and their banner implementations over time.

\subsection{Methodology}

\subsubsection{CMP detection}
\label{sec:method_cmp_detection}
To identify CMPs on visited websites, we utilized Easylist Cookie \cite{easylistcookie}, manually filtering the most frequent URL and CSS rules as established by Bouhoula et al. \cite{bouhoula2024automated}.

We verified CMP ownership by cross-referencing matched domains with official service websites or GitHub implementation documentation. This process identified 18 distinct CMPs.


\subsubsection{CMP types}
Using this method, we detected a total of 18 CMPs. Consistent with prior research \cite{nouwens2025cross,kusk2025website}, we categorize these CMPs into three distinct types. Twelve companies fall under the category of \textit{Comprehensive CMPs}, as they offer a range of services beyond basic consent management, such as website scanning and data insights (e.g., OneTrust, Cookiebot). Two companies are classified as \textit{Plugin CMPs}. These are advertised as lightweight solutions that prioritize easy integration with web applications and frameworks (e.g., Consent Manager, CookieYes). Finally, four companies belong to \textit{Open-source CMPs}, which are designed to be lightweight, freely accessible, and open-source (e.g., Cookie Consent, Cookie Notice).


\subsection{Results}
\subsubsection{CMP popularity}

\begin{figure}[h]
	\includegraphics[alt={The figure shows the top 10 most widely used CMPs and how their use has become more popular over time. Some of the top most popular CMPs have emerged after 2018 or around 2020.}, width=\linewidth]{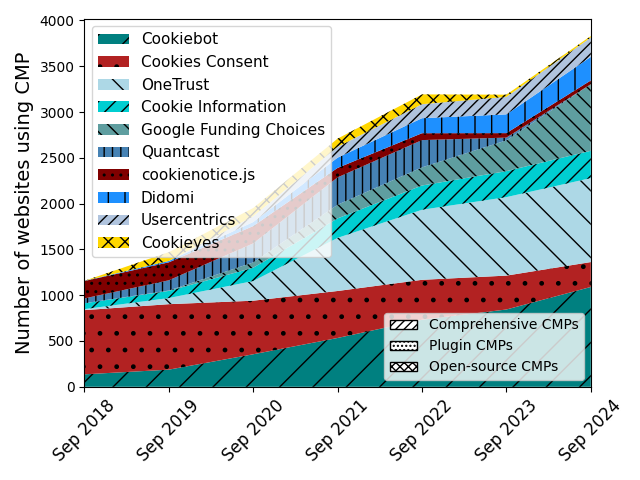}
    \caption{Number of websites which use the top 10 CMPs over time.}
    \label{fig:cmps_popularity}
\end{figure}

We observed a significant increase in CMP adoption over time. The number of websites utilizing CMP banners rose from 1,318 (11.60\%) in 2018 to 4,578 (40.29\%) in 2024. Figure \ref{fig:cmps_popularity} illustrates the yearly evolution of the top 10 most widely used CMPs.

We observe a decline in popularity of \textit{plugin} and \textit{open-source CMPs}. Many were widely used immediately after the GDPR's introduction in 2018. For instance, 53.19\% of websites with a CMP employed the \textit{Cookies Consent} framework \cite{CookiesConsent}, but this figure dropped to just 5.94\% by September 2024. The popularity of all plugin and open-source CMPs has decreased significantly over the last few years.

Conversely, the use of more comprehensive CMPs has grown steadily since 2018. \textit{Cookiebot} \cite{cookiebot} saw its prevalence increase from 136 (10.32\%) websites in 2018 to 1090 (23.81\%) in 2024, potentially because of its merger with \textit{Usercentrics} \cite{UsercentricsMerger}. 
Similarly, the second-largest comprehensive CMP, \textit{OneTrust} \cite{onetrust}, has also seen rapid adoption, increasing from 11 websites (0.83\%) in 2018 to 917 (20.03\%) in 2024.

Google Funding Choices was introduced in 2018 to manage visits from users with ad blockers \cite{gfc2018}. In 2020, a CMP functionality was added \cite{gfc2020}, and it is now integrated with Google's advertising platforms \cite{gfc2024}. In 2024, 727 (15.88\%) websites included resources from Google Funding Choices domains. However, this number represents an upper bound, as it is not possible to distinguish whether these websites are using the CMP functionality or only the anti-adblocking feature without having an actual screenshot of the page.

Overall, our findings suggest a consolidation in the market, with smaller players losing ground to larger, comprehensive CMPs. This indicates that website owners are increasingly willing to invest in more elaborate, and often more expensive, CMP solutions, reflecting a greater concern for (apparent) compliance.

\subsubsection{Cookie banners across CMPs}

To understand how CMPs influence compliance and interface design, we focus on websites utilizing detected CMP-provided banners. We again employ the presence of a reject option at any layer as a proxy for compliance.
In practice, banner deployment is a collaboration between CMPs, which provide a default interface, and website owners, which can modify it at any time. This flexibility makes it difficult to distinguish whether compliance trends stem from the CMP's default settings or the owner’s custom adjustments. To better understand this dynamic between CMPs and publishers, we conduct a series of experiments.

\begin{figure}[h]
	\includegraphics[alt={OneTrust and Cookiebot interfaces have evolved towards having more accept and reject options, but only on about 45\% of OneTrust banners and 55\% of Cookiebot Banners. For Cookies Consent the ratio of accept and reject options remains small. About 30\% of cookie banners which do not use a CMP include accept and reject options, therefore following the compliance trend.}, width=\linewidth]
	{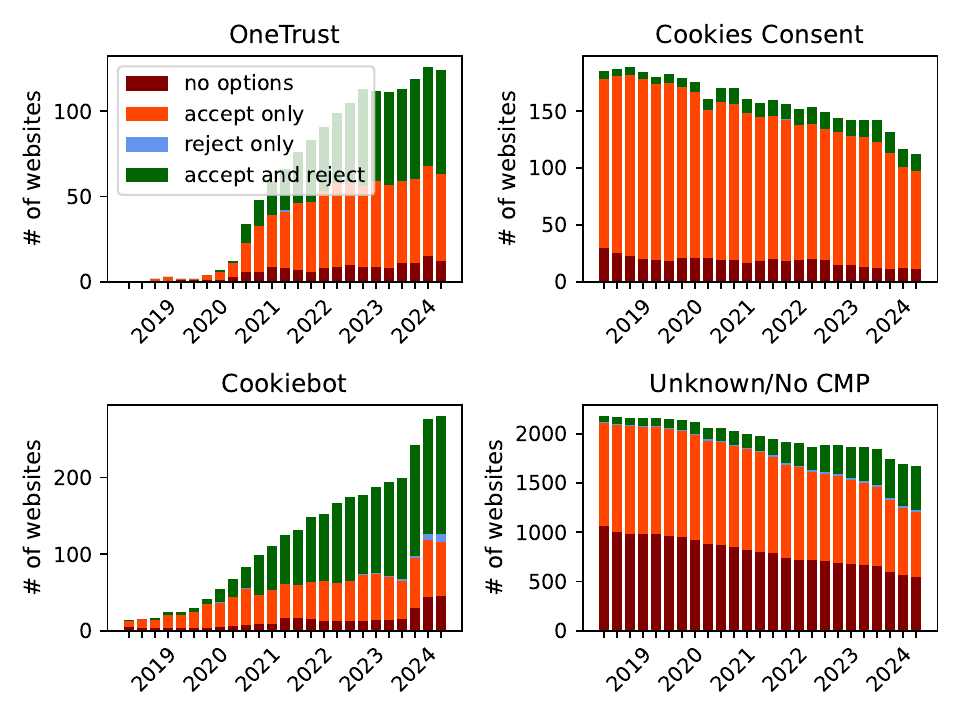}
	\caption{Number of cookie banners with ``Reject'', ``Accept''-only and no options for top 3 CMPs and websites with no CMP}
	\label{fig:options_CMPs}
\end{figure}

\paragraph{What do CMP banners look like?}
In this section, we analyze how CMP-specific banner designs change over time to evaluate whether CMPs influence the prevalence of banners lacking a ``reject'' button. As shown in Figure \ref{fig:options_CMPs}, there are significant differences in compliance rates among various CMPs. For instance, in 2024, Cookiebot demonstrates a compliance rate of 55.16\% and OneTrust 49.19\%, while the rate for Cookies Consent is only 13.39\%. This suggests a varying relationship between the choice of a CMP and the likelihood of having a compliant banner. It seems that the cookie banners of some CMPs are more privacy-friendly than in-house banners, as shown by a 2024 compliance rate of only 26.69\% for non-CMP banners.

The evolution of specific CMPs provides further insight. Cookiebot has steadily evolved toward higher compliance, becoming one of the most widely-used CMPs by 2024. Its most significant shift toward including more reject buttons occurred between June 2020 and December 2021, a period that predates the major global evolution observed in 2022.

In May 2021, noyb filed complaints against 500 OneTrust-using websites for GDPR violations \cite{noybcompliantsmay2021}. By August 2021, 42\% of these sites had remediated their issues \cite{noybcompliantsaugust2021}. This campaign had a measurable effect in our data: the compliance rate for OneTrust banners rose from 33.33\% in March 2021 to 47.47\% in December 2022, following a second wave of enforcement \cite{noybcompliantsaug2022}. Because noyb contacted website owners directly before filing formal complaints, these improvements likely resulted from publishers responding to legal pressure rather than the CMP proactively driving compliance.

\paragraph{What drives positive changes?}
When focusing on the evolution of positive change over time, we observe a similar trend for both CMP and non-CMP banners. This indicates that major regulatory events, such as the CNIL case in March 2022, spurred a significant jump in compliance rates across both groups. This shared, responsive behavior suggests that website publishers actively made changes, instead of CMPs pushing for them. 

Furthermore, we compare existing customers (i.e., websites that were already using the CMP in the previous period) and new CMP customers (i.e., websites that either switched from one CMP to another or did not include a CMP in the previous period). 
We hypothesize that CMPs would actively encourage compliance in both new and existing customers as a result of the CNIL case.  For new customers, this motivation could manifest as updated, more compliant default settings or clearer information about illegal practices. For existing customers, CMPs could either notify them of non-compliant banners or, in more extreme cases, force a change by no longer supporting non-compliant interfaces.
However, our data does not support this hypothesis. As shown in Figure \ref{fig:options_CMPs}, at no point have all interfaces from major CMPs become compliant. If CMPs were indeed a primary driver of compliance, we would expect to see a widespread change, such as 90\% of all websites adopting ``reject'' buttons. 

Consequently, we conclude that CMPs have had a limited impact on this upward trend in compliance, and show little response to regulatory action.

\section{Discussion}

Between 2018 and 2024, compliance with privacy regulations among website owners has increased, likely due to a combination of factors: the threat of fines and regulatory action, concerns over reputational damage, and a general increase in privacy awareness. Although the specific motivator for changes in cookie banner design is difficult to pinpoint, this paper explores the cross-country impact of major actors influencing these banners: website owners, regulators and CMPs.

\subsection{Historical evolution of cookie banners}

Our findings indicate improvements in specific aspects of GDPR compliance. In 2018, only 2.94\% of websites offered both accept and reject options. By 2024, this figure had risen to 30.66\%. This improvement was most significant during two distinct periods. The first major shift occurred in 2022, likely due to a combination of factors, including a wave of complaints filed by noyb with DPAs and the CNIL fine targeting the lack of a reject button on the first layer of the cookie banner.

A second substantial compliance surge occurred in 2024. While difficult to attribute to a single cause, this shift potentially follows the EDPB’s October 2023 guidelines clarifying the ePrivacy Directive. Increased DPA enforcement and rising privacy awareness also likely played key roles. Given this momentum, the trend toward higher compliance will likely persist into 2025 and beyond, providing a significant area for future study.

We acknowledge that our findings should be interpreted as being lower-bound, given that the use of web archives proved to result in a variety of cookie banners being missed by our analysis. Nonetheless, we argue that the methodology allows us to observe GDPR compliance trends over time. 
Overall, our findings show a positive evolution in cookie banner compliance, driven in part by regulatory action (e.g., case law) and  guidance (e.g., EDPB guidelines). While this progress is positive, new privacy issues, such as manipulation and user choices not being respected, are still prevalent.

\subsection{The role of DPAs}

Our results show considerable differences in compliance rates across EU countries. Denmark, France, and Austria have consistently been  part of the top countries with the greatest compliance, with Spain, Finland, and Slovakia catching up in 2024. Conversely, Lithuania, Bulgaria, and Slovenia remain at the lower end of the compliance spectrum.

Higher compliance rates correlate with active DPAs, yet other factors significantly influence banner design. Beyond regulators, NGOs like noyb drive compliance through strategic complaints. Furthermore, a ``spillover effect'' can boost compliance in countries with inactive DPAs, such as Slovakia, while others like Slovenia remain stagnant due to limited enforcement resources. EU-wide EDPB guidelines also exert varying degrees of influence across member states, making it difficult to isolate a single motivation for compliance. Cross-country divergence complicates compliance further with contradictory regulatory guidelines. For example, some DPAs disagree on whether ``accept'' and ``reject'' buttons must be on the same interface level, with the Italian DPA specifically discouriging an equal path for both: ``the affirmative action the user is empowered to perform when first accessing a website must in any case be aimed at giving his or her consent (so-called ``opt-in'') and may never consist in refusing such consent (so-called ``opt-out'').'' \cite{guidelinesItalianDPA}.

Our findings document the effect of enforcement on compliance and highlight the critical importance of DPA collaboration. Many DPAs lack the resources to counter the behavioral advertising industry's evolving manipulative tactics, such as the recent \textit{pay-or-consent} model. Even with adequate resources, the enforcement process can be challenging and span years, especially when well-resourced parties are involved. Additionally, the average website often uses third-party services that operate across multiple EU countries, further complicating enforcement.

We argue that greater EU-wide uniformity is needed for two primary reasons: first, to foster collaboration, enhancing enforcement and support for website owners; and second, to standardize cookie banner implementations. Such uniformity could be realized through Article 88b of the Digital Omnibus proposal \cite{omnibus}. Legally mandating a technically supported consent signal (e.g., via browser vendors) would grant users more control, minimize banner divergence, streamline regulatory oversight, and ultimately increase compliance.

\subsection{The role of CMPs}

\subsubsection{Market share}
Our findings show the growing significance of CMPs as part of the adtech ecosystem. Their adoption has become increasingly popular, rising from 11.6\% of websites in 2018 to 40.29\% in 2024, though this number is likely higher, as smaller CMPs may have been missed by our methodology.
We observe a market trend where website owners are opting for more comprehensive and costly privacy solutions. This shift indicates a greater concern for compliance and a willingness to spend more money on privacy solutions. Consequently, the CMP market is converging around a few popular players, with OneTrust, Cookiebot, and Google Funding Choices holding a combined 61.85\% market share in 2024.
This dominant position of CMPs has been scrutinized by prior work \cite{nouwens2025cross,jha2024privacy} given their presence on popular websites \cite{hils2020measuring}.

\subsubsection{How do CMPs influence cookie banners' design?}

We found significant differences in compliance rates among CMPs. For example, in 2024, 55.16\% of websites using Cookiebot included a reject button, compared to only 13.39\% of sites using Cookies Consent. This contrasts with the 26.69\% rate for websites using their own banner, indicating that the choice of CMP can strongly influence compliance. 
These cross-CMP variations stem from several factors. In some countries, locally developed CMPs are more common, which helps explain some of the cross-country differences we've observed. Furthermore, certain CMPs might use manipulative designs during installation, steering website owners toward less privacy-friendly interfaces \cite{toth2022dark}.

In theory, the evolution of cookie banners on CMP‑using websites reflects both updates made by CMPs and configuration choices made by website owners. In practice, our results indicate that website owners are the primary drivers of change.
First, regulatory actions are typically directed at website owners, not the CMPs they use. Regulations may impact CMPs indirectly (as seen in noyb's complaints against OneTrust), but it is the website owners who are the ultimate agents of change. Second, we see a similar rise in compliance over time on websites that don't  use a CMP, which suggests a broader increase in awareness among owners. Third, CMPs show little response to regulatory actions and do not incentivize customers to adopt more compliant configurations. 

We conclude that the increased compliance is predominantly due to website owners intentionally configuring their banners to meet GDPR standards. This shift is likely fueled by a growing awareness of GDPR violations, driven by significant enforcement actions (e.g., noyb complaints, the CNIL case) and a general rise in public concern over privacy and reputational risk.

\subsubsection{Responsibility and compliance}

Our study reveals a positive trend in cookie banner compliance over time. However, persistent issues like manipulative designs and disregarded user choices raise a fundamental question: who is responsible for compliance? By examining various stakeholders, our results show that most improvements stem from website owners actively updating their banners. In contrast, despite their significant market share and close ties to the adtech ecosystem, CMPs exert limited effort to improve compliance. Because they position themselves as expert solutions, they effectively set the de facto standard for cookie banners, leading developers to rely heavily on their default configurations. Furthermore, prior work shows that many website owners are unaware of the various technologies on their pages and rely on tools such as cookie scans for compliance \cite{kusk2025website}. This issue is compounded by the fact that some CMPs include manipulative designs in their configuration process for new customers~\cite{toth2022dark} and offer tools like ``consent optimization'' and A/B testing of interface configurations that may not serve users' best interests. 
CMPs are likely to have more incentive to maximize user consent, as it might give them a competitive edge with relation to other market players and help to retain clients.
At the very least, this behavior is misleading to the publishers who rely on their services.

At the same time, website owners retain their own responsibilities. One could argue that CMPs merely provide the tools, and website owners must still configure them correctly to meet GDPR requirements. Some CMPs might be more inclined to meet those criteria bu default, in order to guarantee safety from enforcement actions for their customers. 

In practice, we argue that GDPR compliance for cookie banners is a shared responsibility between CMPs and publishers. There is a clear need for greater legal accountability for CMPs. Revisiting questions around joint controllership could help determine when CMPs should be held responsible under the GDPR \cite{santos2021consent}. At present, CMPs have little incentive to push for higher compliance and show minimal response to regulatory actions. Stronger incentives could help reduce manipulative designs and lead to more efficient improvements in compliance. Additionally, standardizing how users communicate consent would constrain CMPs, leaving less room for non-compliant banner interfaces such as those that lack a clear ``Reject'' option.

Ultimately, increased regulatory oversight of CMPs would more effectively scale privacy protections across the web. Based on our findings, we encourage website owners to be cautious when relying on CMPs' interfaces and instead, ensure privacy-friendly cookie banners which align with users' choices regarding their privacy on the web. 

\section{Conclusion}

Our findings reveal a complex, evolving landscape of cookie banner compliance under the GDPR. Compliance rose from 2.94\% in 2018 to 30.66\% in 2024, driven by the interplay between publishers, CMPs, and DPAs. A strong correlation between DPA activity and higher compliance, particularly in France and Spain, underscores the necessity of consistent enforcement. Conversely, countries with limited regulatory activity, such as Lithuania, Poland, and Bulgaria, maintain notably low compliance.

CMPs offer widely varying interfaces; for instance, ``reject'' options range from 55.16\% on Cookiebot to only 13.39\% on Cookies Consent. Despite their prevalence, CMPs have shown little response to regulatory pressure. Instead, compliance improvements appear driven primarily by publishers opting for more privacy-respecting designs. By analyzing these stakeholders, this study clarifies the forces shaping GDPR compliance and the gradual shift toward privacy-friendly interfaces.

\section*{Acknowledgments}
This research is partially funded by the Internal Funds KU Leuven, and by the Cybersecurity Research Program Flanders.

%
%
%
\bibliographystyle{splncs04}
\bibliography{bib}

\end{document}